\documentclass{article}
\pdfoutput=1
\usepackage{graphicx}
\usepackage{float}
\pdfoutput=1
\usepackage{subcaption}
\usepackage{booktabs} 
\usepackage{multirow}
\usepackage{stfloats}
\usepackage{amsmath}
\usepackage{amssymb}
\usepackage{xcolor}
\usepackage{makecell}
\usepackage{array}
\newcommand{\PreserveBackslash}[1]{\let\temp=\\#1\let\\=\temp}
\newcolumntype{C}[1]{>{\PreserveBackslash\centering}p{#1}}
\newcolumntype{R}[1]{>{\PreserveBackslash\raggedleft}p{#1}}
\newcolumntype{L}[1]{>{\PreserveBackslash\raggedright}p{#1}}
\usepackage{caption}
\usepackage[capitalize]{cleveref}
\usepackage{cite}
\usepackage{spconf,amsmath,graphicx}

\usepackage{enumitem}
\title{Ripple Sparse Self-Attention For Monaural Speech Enhancement}
\name{Qiquan Zhang{$^{1,2}$}, Hongxu Zhu{$^{2,*}$}, Qi Song$^{3}$, Xinyuan Qian{$^2$}, Zhaoheng Ni{$^4$}, Haizhou Li$^{2,5}$
\thanks{Qiquan Zhang, Hongxu Zhu, Xinyuan Qian, and Haizhou Li are supported by 1) National Natural Science Foundation of China (Grant No. 62271432), Internal Project Fund from Shenzhen Research Institute of Big Data (Grant No. T00120220002); 2) Agency for Science, Technology and Research (A*STAR) under its AME Programmatic Funding Scheme (Project No. A18A2b0046); 3) Guangdong Provincial Key Laboratory of Big Data Computing, The Chinese University of Hong Kong, Shenzhen (Grant No. B10120210117-KP02); 4) ARC Discovery Grant DP1900102479; 5) IAF, A*STAR, SOITEC, NXP and National University of Singapore under FD-fAbrICS: Joint Lab for FD-SOI Always-on Intelligent \& Connected Systems (Award I2001E0053).} \thanks{*Corresponding author}}
\address{$^1$School of Electrical Engineering and Telecommunications, University of New South Wales, Australia\\
$^2$Department of Electrical and Computer Engineering, National University of Singapore, Singapore\\
$^3$Alibaba Group, China \, $^4$Meta, United States \\
$^5$ Shenzhen Research Institute of Big Data, School of Data Science, \\ 
The Chinese University of Hong Kong, Shenzhen, China}

\begin{document}
\ninept
\maketitle
\vspace{-0.5em}
\begin{abstract}
The use of Transformer represents a recent success in speech enhancement. However, as its core component, self-attention suffers from quadratic complexity, which is computationally prohibited for long speech recordings. Moreover, it allows each time frame to attend to all time frames, neglecting the strong local correlations of speech signals. This study presents a simple yet effective sparse self-attention for speech enhancement, called ripple attention, which simultaneously performs fine- and coarse-grained modeling for local and global dependencies, respectively. Specifically, we employ local band attention to enable each frame to attend to its closest neighbor frames in a window at fine granularity, while employing dilated attention outside the window to model the global dependencies at a coarse granularity. We evaluate the efficacy of our ripple attention for speech enhancement on two commonly used training objectives. Extensive experimental results consistently confirm the superior performance of the ripple attention design over standard full self-attention, blockwise attention, and dual-path attention (SepFormer) in terms of speech quality and intelligibility.

\end{abstract}
\section{Introduction}
\label{sec:intro}

The perceived quality and intelligibility of speech signals are inevitably degraded by surrounding background noises in our daily acoustic scenarios. As a solution, speech enhancement improves the quality of speech perception by separating clean speech from degraded noisy speech. Due to some underlying assumptions, traditional speech enhancement methods often lack the ability to suppress non-stationary noises \cite{mmse,gerkman2013,zhang2019,8740919}. Over the past decade, deep learning has promoted tremendous progress in speech enhancement given the supervision of large-scale training data \cite{wang2018supervised}. A typical method is to optimize a deep neural network (DNN) to predict the spectra of clean speech or a time-frequency (T-F) mask from the T-F spectra of noisy speech. Motivated by the idea of T-F auditory masking, the ideal binary mask (IBM) is introduced for speech enhancement \cite{wang2018supervised}. Subsequently, various T-F masks are proposed, including ideal ratio mask (IRM) \cite{wang2014training}, complex IRM (cIRM) \cite{williamson2015complex}, and phase-sensitive mask (PSM) \cite{erdogan2015phase}. DNNs are also optimized to recover the waveform of clean speech directly from the noisy raw waveform in an end-to-end manner \cite{SEGAN,realse}.

With the ability to model the long-range correlations, long short-term memory networks (LSTMs) are adopted for speech enhancement \cite{weninger2015speech,erdogan2015phase}. Nevertheless, LSTM network architectures involve a large number of parameters and inherent sequential nature, which excludes its use for many applications. Convolutional neural networks process time frames in parallel and can also capture contextual dependencies by stacking multiple layers. Residual temporal convolution networks (ResTCNs) \cite{TCN2018}, incorporating 1-D dilated convolution and residual connection, have been applied for speech enhancement with an impressive performance \cite{deepmmse,GRN,zhang2021temporal,TFA}.


Transformer \cite{transformer} has recently shown state-of-the-art (SOTA) results in numerous speech tasks, including speech enhancement \cite{mhanet,tgsa,sepformerstft,tfaj}. As a core piece of Transformer, the multi-head self-attention attends to all time steps in parallel, allowing Transformer to model long-term dependencies efficiently. However, the computation complexity of self-attention is quadratic with respect to the sequence length, thus making it infeasible for long speech recordings. In addition, the strong local correlations of speech are ignored. The more recent dual-path transformer (SepFormer) \cite{sepformer} splits the sequence into overlapped chunks and conducts self-attention to intra and inter chunks, for local and global modeling, respectively, which mitigate the issue of quadratic complexity. \textcolor{black}{SepFormer is introduced in \cite{sepformer} for speech separation, demonstrating SOTA performance, and then applied for speech enhancement \cite{sepformerstft}.} \textcolor{black}{We present a sparse attention for speech enhancement, here referred to as ripple attention, which can simultaneously model the local and global correlations in a spectrogram in a simpler and more effective way, without chunking.} In particular, the ripple attention consists of fine-grained local attention and coarse-grained global attention. Considering that speech frames usually have a stronger correlation in a local context than in a broad context over time, the local attention restricts each frame to attend to its neighboring frames in a context window at fine granularity, which takes special care of neighboring frames. The dilated attention is applied to the time frames outside the window at a coarse granularity, allowing the model to capture the global dependencies more efficiently. \textcolor{black}{In this paper, we show that our ripple attention outperforms blockwise attention and the SOTA SepFormer on two commonly used training targets, and theoretically analyze the computation cost.} The closest work to ours is Longformer \cite{longformer} for document processing, where the dilated attention is restricted in a sliding window for local modeling, unlike our model, which applies dilated attention outside the window, for coarse-grained global modeling. Moreover, our speech-oriented attention allows each frame for global modeling instead of only a few task-specific tokens allowed to query all tokens for global modeling. 

The rest of this paper is organized as follows. In Section \ref{sec:2}, we introduce T-F neural speech enhancement. Section \ref{sec:3} describes our proposed method. Section \ref{sec:4} sets up the experiment and analysis the results. Section \ref{sec:5} gives the conclusion.

\section{Problem Formulation}\label{sec:2}

Taking the short-time Fourier transform (STFT), the noisy waveform is transformed into T-F domain:
$X[l,k]\!=\!S[l,k]+D[l,k]$,
where $l$ is the time frame index and $k$ is the frequency bin index. $S[l,k]$, $D[l,k]$ and $X[l,k]$ represent the coefficients of clean speech, noise, and noisy speech, respectively. A typical T-F neural speech enhancement optimizes a DNN to predict a T-F mask, $\widehat{M}[l,k]$, to separate the clean speech, $\widehat{S}[l,k]\!=\widehat{M}[l,k]\cdot\!X[l,k]$. \textcolor{black}{Without the loss of generality, we adopt two commonly used T-F masks, IRM \cite{wang2014training} and PSM \cite{erdogan2015phase}, to validate our proposed sparse self-attention mechanism.}
\begin{equation}\label{IRM}
\setlength{\belowdisplayskip}{2.5pt}
\text{IRM}[l,k]=\sqrt{\frac{|S[l, k]|^{2}}{|S[l, k]|^{2}+|D[l, k]|^{2}}}
\end{equation}
where $|\cdot|$ denotes the spectral magnitude. 
\begin{equation}\label{PSM}
\text{PSM}[l,k]=\frac{\left|S[l, k]\right|}{|X[l, k]|}\cos[\theta_{S[l, k]-X[l, k]}]
\end{equation}
where $\theta_{S[l, k]-X[l, k]}$ indicates the phase difference between the STFT coefficients of clean and noisy speech \cite{erdogan2015phase}.

\section{Speech Enhancement with Ripple Sparse Self-Attention}\label{sec:3}
\subsection{Network Architecture}
\begin{figure}[!ht]
\centering
\begin{subfigure}[t]{0.33\columnwidth}
\centerline{\includegraphics[width=0.95\columnwidth]{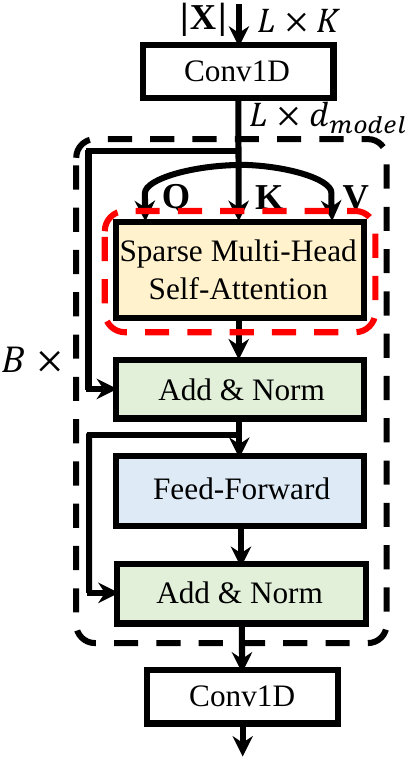}}
\caption{}
\label{fig1:1}
\end{subfigure}
\begin{subfigure}[t]{0.62\columnwidth}
\centerline{\includegraphics[width=1.0\columnwidth]{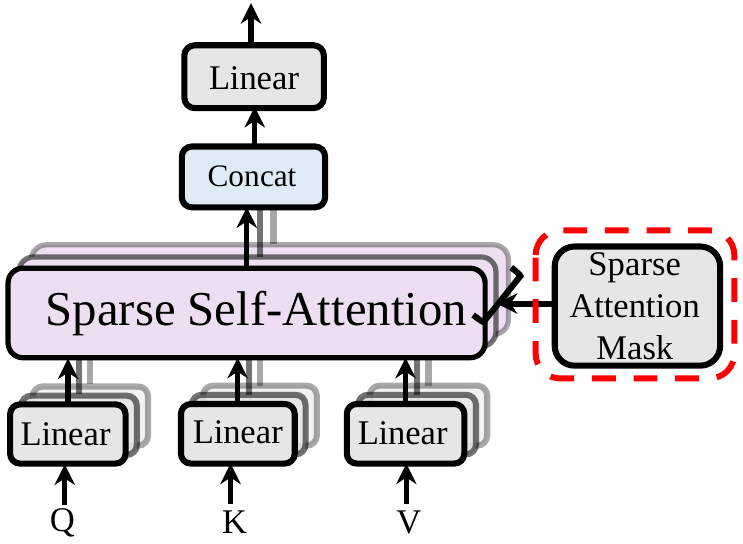}}
\caption{}
\label{fig1:2}
\end{subfigure}
\vspace{-0.8em}
\caption{Illustration of (a) the ripple sparse self-attention backbone network and (b) the sparse multi-head self-attention module. The key idea is to introduce a sparse attention mask to sparsify $L\!\times\!L$ attention matrix.}
\label{fig1}
\vspace{-1.5em}
\end{figure}

Fig.\ref{fig1} (a) illustrates the overall architecture of our sparse self-attention backbone network. The STFT magnitude spectrum of noisy speech $|\textbf{X}|\!\in\!\mathbb{R}^{L\times K}$ is the input to the network, where $L$ is the number of time frames and $K$ is the number of frequency bins. The input is first projected by Conv1D, a 1-D convolution layer of kernel size 1 that involves a frame-wise layer normalization with the ReLU as activation function, which produces a latent T-F representation with a size of $d_{model}$. The produced latent representation is then fed into $B$ stacked sparse self-attention blocks, and each block consists of two subblocks: a sparse multi-head attention (MHA) module and a two-layer fully connected feed-forward network (FNN). A residual connection is applied around each subblock, followed by frame-wise layer normalization. The output layer of the network is a 1-D convolution layer of kernel size 1 and the sigmoid activation function is applied to output the estimated mask.

\subsection{Sparse Self-Attention}
Fig.\ref{fig1} (b) illustrates the workflow of the ripple sparse multi-head attention, which takes as input a set of queries ($\textbf{Q}\!\in \!\mathbb{R}^{L\times d_{model}}$), keys ($\textbf{K}\!\in\!\mathbb{R}^{L\times d_{model}}$), and values ($\textbf{V}\!\in\!\mathbb{R}^{L\times d_{model}}$). 
A total of $h$ attention heads are employed in the sparse multi-head attention, where the head index is denoted as $i\!=\!\left\{1,2,3,...,h\right\}$, allowing the model to pay attention to different aspects of information. 
\textcolor{black}{For $i$-th head, firstly a linear projection is applied to $\textbf{Q}$, $\textbf{K}$, and $\textbf{V}$, respectively: $Q_{i}\!=\!\textbf{Q}\textbf{W}_{i}^{Q}$, $K_{i}\!=\!\textbf{K}\textbf{W}_{i}^{K}$, and $V_{i}\!=\!\textbf{V}\textbf{W}_{i}^{V}$, with dimensions $d_{k}$, $d_{k}$, and $d_{v}$, respectively, where $\textbf{W}_{i}^{Q}\in \!\mathbb{R}^{d_{model}\times d_{k}}$, $\textbf{W}_{i}^{K}\in\!\mathbb{R}^{d_{model}\times d_{k}}$, and $\textbf{W}_{i}^{V}\in\!\mathbb{R}^{d_{model}\times d_{v}}$ are learned parameter matrices.}
For each head, the scaled dot-product attention is utilized to compute the attention scores. The output is given as:
\begin{equation}
    A_{i} = \text{softmax}\left(\frac{Q_{i}K_{i}^\top}{\sqrt{d_{k}}}\odot M \right)V_{i}
\end{equation}
where $d_{k}=d_{v} = d_{model}/{h}$, $M\!\in\left\{0,1\right\}^{L \times L}$ denotes the attention mask used to sparsify the attention matrix, where $M_{i j}=1$ indicates that frame $i$ is allowed to attend to frame $j$, 0 otherwise. The operation $\odot$ is defined as: 
\begin{equation}
\setlength{\belowdisplayskip}{5pt}
 \left(P \odot M\right)_{i j}= \begin{cases}P_{i j} & \text { if } M_{i j}=1 \\ -\infty & \text { if } M_{i j}=0\end{cases}
\end{equation}
The outputs for each attention head are then concatenated and linearly projected:
\begin{equation}
    \text{MultiHead}\left(\textbf{Q}, \textbf{K}, \textbf{V}\right) = \text{Concat}(A_{1},...,A_{h})\textbf{W}^{O}
\end{equation}
where $\textbf{W}^{O}\!\in\!\mathbb{R}^{d_{model}\times d_{model}}$. \textcolor{black}{The two-layer FFN takes the output of the first subblock and performs two linear transformations with a ReLU activation function in the first layer. Detailed descriptions of our ripple sparse attention pattern are given in the next section.}


\begin{figure}[!hbtp]
\centering
\begin{subfigure}[t]{0.30\columnwidth}
\centerline{\includegraphics[width=1\columnwidth]{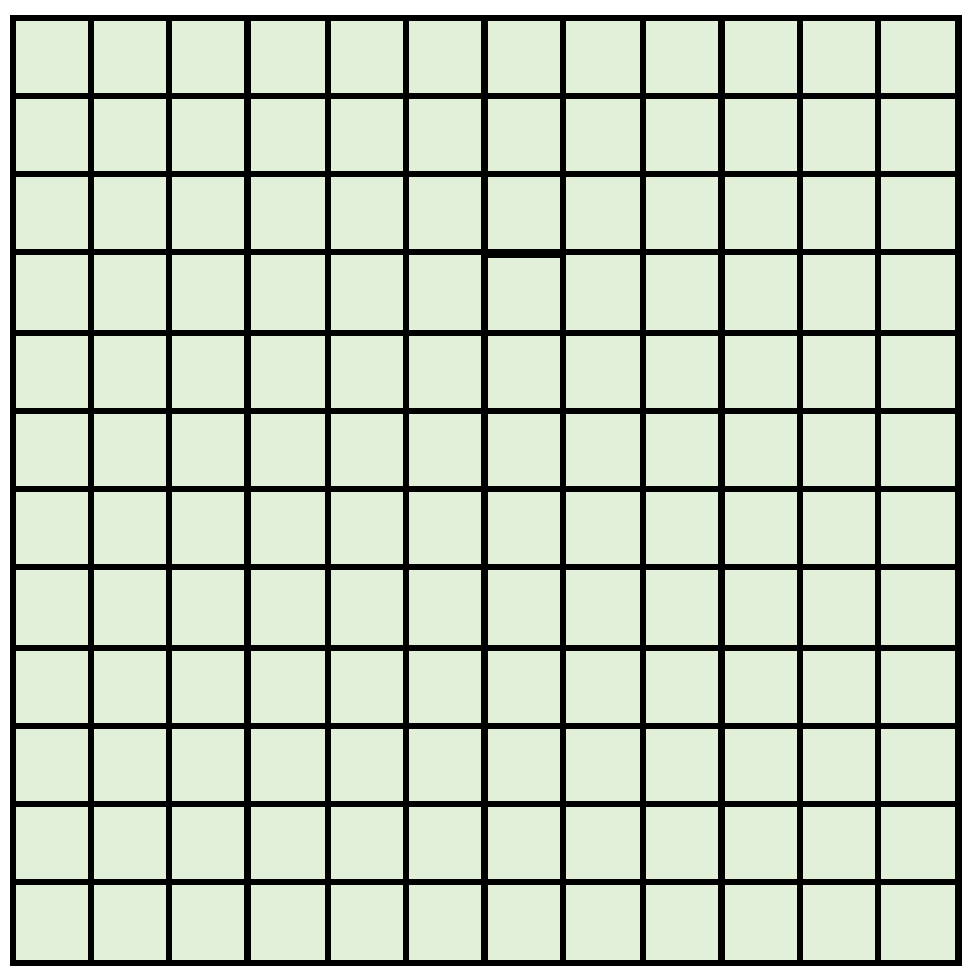}}
\caption{Full self-attention}
\label{fig2:1}
\end{subfigure}
\hspace{4mm}
\begin{subfigure}[t]{0.31\columnwidth}
\centerline{\includegraphics[width=0.955\columnwidth]{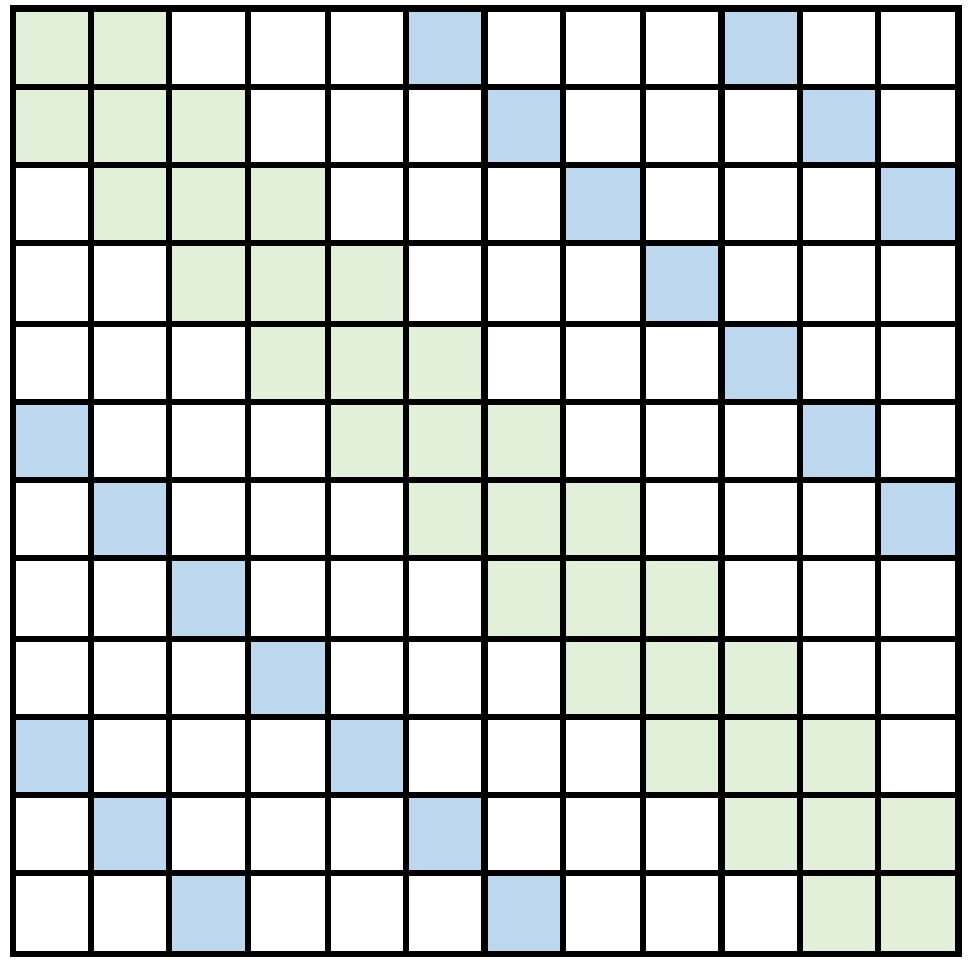}}
\caption{Our proposed ripple sparse self-attention}
\label{fig2:2}
\end{subfigure}
\caption{Illustration of the full self-attention pattern and the ripple sparse attention pattern (with sequence length $L=12$).}
\vspace{-1.5em}
\label{fig2}
\end{figure}
\subsection{Ripple Sparse Attention Mask}

Fig. \ref{fig2} shows an example of the full-attention mask and our ripple sparse-attention mask. The colored squares (with the row index $i$ and column index $j$) correspond to the mask value $M_{ij}\!=\!1$, which indicates that frame $i$ attends to frame $j$ and the corresponding attention score is involved. The blank squares correspond to $M_{ij}=0$, which indicates that the attention score is masked or discarded. As illustrated in Fig. \ref{fig2} (a), full attention computes the correlations between all time frames, which leads to $O(L^{2})$ complexity that limits its use for long audio recordings. In addition, it also inhibits local correlations among nearby frames. To tackle these issues, we propose a ripple sparse attention that involves fine-grained local and coarse-grained global modeling, enabling the model to capture both local and global dependencies more efficiently and effectively. 


Since the speech signal comes with strong local correlations, our ripple attention employs a fine-grained local attention that controls each frame to attend to the frames within the context window. Specifically, given a sliding window with a length of $w$ frames, each time frame only attends to $\frac{1}{2}w$ frames on the left and right sides (green squares in Fig. \ref{fig2} (b)), which achieves linear complexity $O(Lw)$. To allow the model to capture the global dependencies, our attention employs dilated self-attention for the frames outside the local context window at a coarse granularity. The dilated self-attention is only computed between positions that are $d$ time frames away from each other, which has $O(L\lfloor\frac{L-w}{d}\rfloor)$ complexity. This is analogous to dilated CNNs where the window has dilation rate gaps $d$ (blue squares in Fig. \ref{fig2} (b)).

\section{EXPERIMENTS}
\label{sec:4}
\subsection{Datasets and Feature Extraction}

\textcolor{black}{Following the datasets setup in \cite{tfaj}, we employ the Librispeech \cite{panayotov2015librispeech} \textit{train-clean-100} set ($28\,539$ utterances) as the clean speech data in the training set. The noise data contains a total of $6\,909$ noise recordings, which are from seven datasets: the coloured noise recordings \cite{mhanet}, the noise set in the MUSAN corpus \cite{snyder2015musan}, the Nonspeech noises \cite{hu2010tandem}, the QUT-NOISE dataset \cite{dean2010qut}, the Environmental Noise dataset \cite{saki2016smartphone, saki2016automatic}, the RSG-10 dataset \cite{steeneken1988description}, and the Urban Sound dataset \cite{Urban}. We randomly exclude 1000 clean speech utterances and noise recordings to yield 1000 noisy/clean pairs as the validation set, where a segment randomly selected from one noise recording is mixed with one clean speech at an SNR value (randomly chosen from $[-10,\,20]$ dB, in 1 dB increments). Four real-world noises (\textit{babble voice}, \textit{F16}, and~\textit{factory welding} excluded from RSG-10 dataset \cite{steeneken1988description} and \textit{street music} excluded from the Urban Sound dataset \cite{Urban}) are used for testing. We randomly chose ten clean speech utterances from the Librispeech \textit{test-clean-100} set for each noise recording, and mix each clean speech utterance with a noise segment randomly selected from the noise recording at SNRs of -5, 0, 5, 10 and 15 dB, which yields 200 noisy speech for testing. All audio signals in the experiment are sampled at 16 kHz. For the spectral analysis, we employ a square-root-Hann window of length 32 ms (50\% overlapping) and a 512-point STFT, which leads to the 257-point STFT magnitude spectrum.
}

\subsection{Experimental Setup}
\textcolor{black}{As baselines, we compare our ripple attention with standard full self-attention \cite{transformer,mhanet,restcnsa,tgsa} and two recent sparse attention patterns, blockwise attention \cite{blockwise} (for document processing) and SepFormer (SOTA efficient self-attention model for speech enhancement) \cite{sepformer,sepformerstft} to demonstrate its effectiveness and superiority. All models employ $B\!=\!4$ transformer layers (same model size) and adopt the following parameters: $H\!=\!8$, $d_{model}\!=\!256$, and $d_{f\!f}\!=\!1024$. Self-attention was found to tend to model short-term dependencies in the lower layers \cite{localmodeling}. Inspired by this, \textcolor{black}{unlike \cite{tii} as well as baseline models, the proposed model applies only local band attention to the first two layers.} We investigate the performance of our ripple attention under different configurations for the window size $w$ and the dilation rate $d$: $w\!=\!12$ and $d\!=\!\left\{16,24,50\right\}$. Blockwise attention splits the input sequence into nonoverlapping blocks of size 50 and processes each block separately. \textcolor{black}{For SepFormer baseline \cite{sepformer}, the intra- and inter-chunk transformer blocks include two transformer layers, respectively, and the size of the chunk (or block) is set to 50 (with 50\% overlap) as suggested in \cite{sepformerstft}.}}

\textcolor{black}{Ten clean speech utterances are included in one mini-batch for one training step, and the noisy mixture is generated by mixing each clean speech utterance in the mini-batch with one segment randomly selected from one randomly chosen noise recording at an SNR value randomly chosen integer from the range of $[-10,\,20]$ dB. The loss function is the mean squared error between the ground truth and the predicted mask. For each epoch, we randomly shuffle the order of the clean speech utterances. The \textit{Adam} optimizer with parameters as in \cite{transformer}, i.e., $\!\beta_{1}\!=\!0.9$, $\!\beta_{2}\!=\!0.98$, and $\epsilon\!=\!10^{-9}$ is adopted for training. The gradient clipping is adopted to clip \textcolor{black}{gradient values} to between $[-1,1]$. As the training of transformer is sensitive to the learning rate \cite{mhanet,transformer}, we follow the warm-up training strategy \cite{transformer}, where the learning rate is adjusted according to the rule:
$lr = d_{model}^{-0.5}\cdot \textrm{min} \left(n\_step^{-0.5}, n\_step \cdot wup\_steps^{-1.5}\right)$,
where $n\_step$ and $wup\_steps$ denote the number of training steps and warm-up training steps, respectively. Following \cite{mhanet}, $wup\_steps\!=\!40\,000$ is adopted for warm-up training. \textcolor{black}{It should be noted that compared to the training strategy in \cite{sepformerstft}, we find that the warm-up training strategy shows better results in our dataset.}}


\subsection{Training \& Validation Loss}

\begin{figure}[!htbp]
\centering
\begin{subfigure}[t]{0.46\columnwidth}\centerline{\includegraphics[width=\columnwidth]{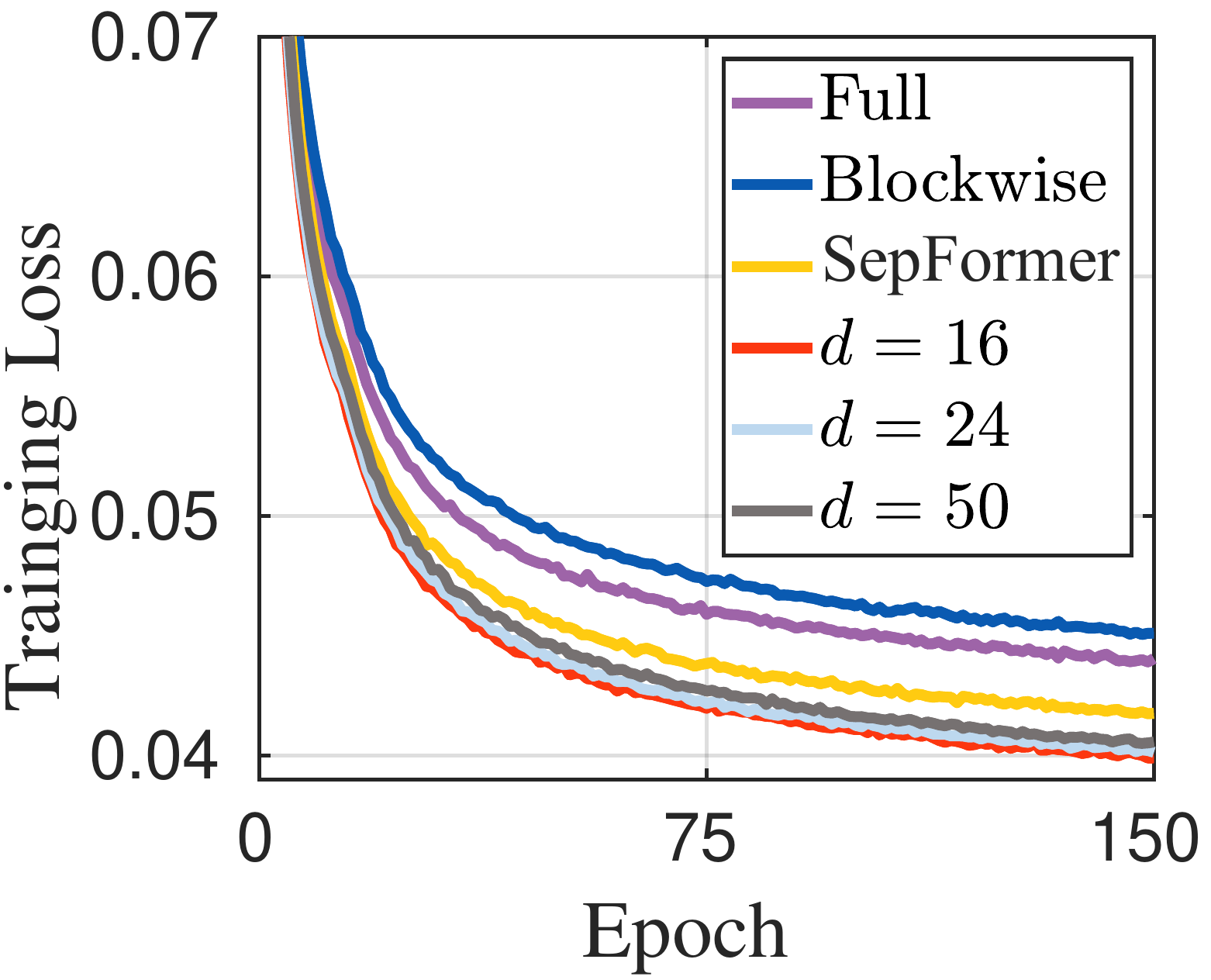}}
\caption{}
\label{fig3:1}
\end{subfigure}
\begin{subfigure}[t]{0.46\columnwidth}
\centerline{\includegraphics[width=\columnwidth]{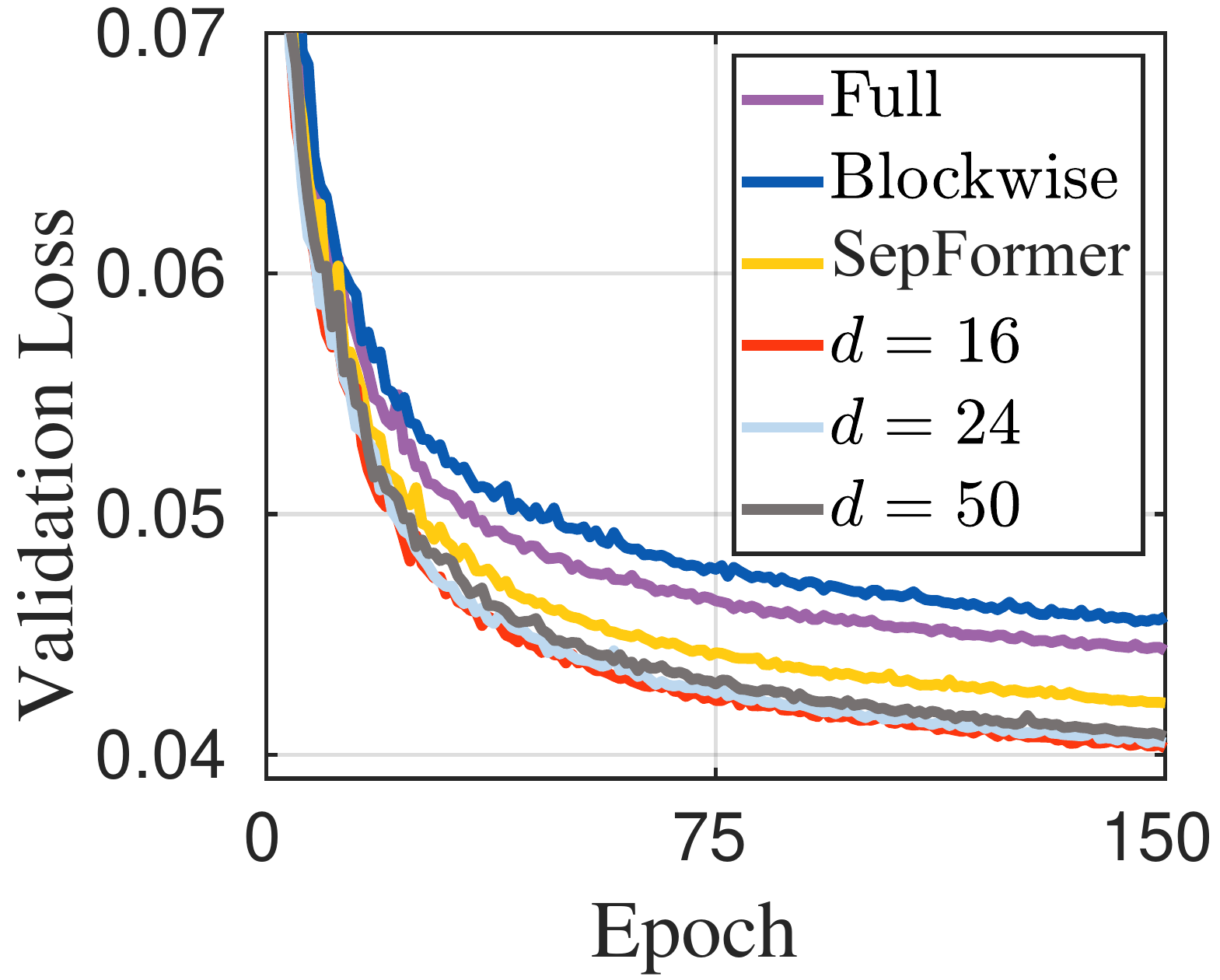}}
\caption{}
\label{fig3:2}
\end{subfigure}
\caption{The (a) training and (b) validation loss curves.}
\label{fig3}
\vspace{-1.0em}
\end{figure}

\textcolor{black}{The training and validation loss curves of different models are illustrated in Fig.~\ref{fig3}, where each of the models is trained for 150 epochs with IRM as the training objective. Compared to standard full self-attention, our ripple attention patterns (with different dilation rates) consistently produce obvious lower training and validation loss, indicating the effectiveness of the ripple attention design. The lower loss than the blockwise attention and SepFormer further demonstrates the superiority of our attention pattern.
}

\begin{table*}[!t]
    \centering
    \footnotesize
    \def\arraystretch{0.95}
    \setlength{\tabcolsep}{4.9pt}
    \caption{Comparison results in terms of PESQ. Boldface scores indicate the best PESQ results for each SNR level.}
    \label{tabwer}
    \vspace*{0.1in}
    \scalebox{0.92}{\begin{tabular}{c|c|c|c|ccccc|ccccc}
        \hline
        \hline
        \multirow{3}{*}{\makecell[c]{Attention\\ Pattern}} & 
        \multirow{3}{*}{\makecell[c]{Block \\ Size}} &
        \multirow{3}{*}{\makecell[c]{Window \\ Size $(w)$}} & 
        \multirow{3}{*}{\makecell[c]{Dilation \\ Rate $(d)$}}
        & & & IRM & & & & & PSM & & \\
        \cline{5-14}
        & & & & \multicolumn{5}{c|}{Input SNR (dB)} 
            & \multicolumn{5}{c}{Input SNR (dB)} \\
        \cline{5-14}
        &  &  &  & -5 & 0 & 5 & 10 & 15 
                 & -5 & 0 & 5 & 10 & 15 \\
        \hline
        \hline
        Noisy & \multirow{2}{*}{--} & \multirow{4}{*}{--} & \multirow{4}{*}{--}
        & 1.28 & 1.50 & 1.84 & 2.21 & 2.58
        & -- & -- & -- & -- & -- \\
        \cline{1-1} 
        \cline{5-14}
        Full Attention &  &  &   & 1.66 & 2.13 & 2.51 & 2.84 & 3.14
                       & 1.76 & 2.25 & 2.65 & 3.00 & 3.32\\
        \cline{1-2}
        \cline{5-14}
        Blockwise & \multirow{2}{*}{50} &  &  
        & 1.65 & 2.11 & 2.51 & 2.84 & 3.15
        & 1.75 & 2.25 & 2.67 & 3.01 & 3.33\\
        \cline{1-1}
        \cline{5-14}
        SepFormer &  &  &  & 
        1.71 & 2.16 & 2.54 & 2.88 & 3.18 & 
        1.78 & 2.30 & 2.69 & 3.04 & 3.34\\
        \hline
        \hline
        \multirow{3}{*}{\makecell[c]{Ripple}}
        & \multirow{3}{*}{--}
        & \multirow{3}{*}{12} & 16 
        & \textbf{1.80} & \textbf{2.26} & \textbf{2.65} & \textbf{2.96} & \textbf{3.27} 
        & 1.87 & 2.37 & 2.79 & 3.13 & \textbf{3.44} \\
               
        &  &  & 24 
        & 1.78 & 2.25 & 2.63 & 2.95 & 3.24  
        & \textbf{1.89} & \textbf{2.39} & \textbf{2.80} & \textbf{3.14} & \textbf{3.44} \\
        &  &  & 50 
        & \textbf{1.80} & \textbf{2.26} & 2.63 & 2.95 & 3.23  
        & 1.85 & 2.36 & 2.78 & 3.13 & 3.42 \\
        \hline
        \hline
    \end{tabular}}
    \label{pesq}
    \vspace{-0.8em}
\end{table*}

\begin{table*}[!t]
    \centering
    \footnotesize
    \def\arraystretch{1.0}
    \setlength{\tabcolsep}{4.9pt}
    \caption{Comparison results in terms of ESTOI (in \%). Boldface scores indicate the best ESTOI results for each SNR level.}
    \label{estoi}
    \vspace*{0.1in}
    \scalebox{0.92}{\begin{tabular}{c|c|c|c|ccccc|ccccc}
        \hline
        \hline
        \multirow{3}{*}{\makecell[c]{Attention \\ Pattern}} & 
        \multirow{3}{*}{\makecell[c]{Block \\ Size}} &
        \multirow{3}{*}{\makecell[c]{Window \\ Size $(w)$}} & 
        \multirow{3}{*}{\makecell[c]{Dilation \\ Rate $(d)$}}
        & & & IRM & & & & & PSM & & \\
        \cline{5-14}
        & & & & \multicolumn{5}{c|}{Input SNR (dB)} 
            & \multicolumn{5}{c}{Input SNR (dB)} \\
        \cline{5-14}
        &  &  &  & -5 & 0 & 5 & 10 & 15 
                 & -5 & 0 & 5 & 10 & 15 \\
        \hline
        \hline
        Noisy & \multirow{2}{*}{--} & \multirow{4}{*}{--} & \multirow{4}{*}{--}
        & 27.91 & 42.14 & 57.21 & 71.11 & 82.22
        & -- & -- & -- & -- & -- \\
        \cline{1-1} 
        \cline{5-14}
        Full Attention &  &  &   
        & 42.11 & 59.86 & 73.87 & 83.70 & 89.88
        & 42.25 & 60.32 & 74.43 & 84.13 & 90.11\\
        \cline{1-2}
        \cline{5-14}
        Blockwise & \multirow{2}{*}{50} &  &  
        & 41.26 & 59.45 & 73.73 & 83.54 & 89.78
        & 41.66 & 60.03 & 74.34 & 83.96 & 90.09 \\
        \cline{1-1}
        \cline{5-14}
        SepFormer &  &  &  
        & 42.58 & 60.19 & 74.25 & 83.87 & 89.99
        & 42.75 & 60.71 & 74.73 & 84.29 & 90.26 \\
        \hline
        \hline
        \multirow{3}{*}{\makecell[c]{Ripple}}
        & \multirow{3}{*}{--}
        
        & \multirow{3}{*}{12} 
          & 16 & 44.89 & \textbf{62.90} & \textbf{76.24} & \textbf{84.85} & 90.43 
               & 45.21 & \textbf{63.43} & 76.78 & 85.27 & 90.76 \\
               
        &  &  & 24 
        & 44.67 & 62.50 & 75.89 & 84.80 & \textbf{90.53}   
        & \textbf{45.24} & 63.40 & \textbf{76.79} & \textbf{85.30} & \textbf{90.84} \\
        
        &  &  & 50 
        & \textbf{44.92} & 62.76 & 76.21 & 84.79 & 90.41 
        & 44.21 & 62.80 & 76.52 & 85.23 & 90.72 \\
        \hline
        \hline
    \end{tabular}}
    \label{estoi}
    \vspace{-1.5em}
\end{table*}

\vspace{-0.8em}
\subsection{Results and Discussion}

\textcolor{black}{The PESQ \cite{recommendation2001perceptual} and ESTOI \cite{jensen2016algorithm} are adopted to evaluate the perceptual quality and intelligibility of speech, respectively. Three composite metrics \cite{hu2007evaluation} are used to evaluate the mean opinion scores of overall signal quality (COVL), signal distortion (CSIG), and background-noise intrusiveness (CBAK).}
\vspace{-1.0em}

\begin{table}[!htbp]
\centering
    \scriptsize
    \def\arraystretch{1.05}
    \setlength{\tabcolsep}{2.5pt}
\caption{Comparisons of CSIG, CBAK, and COVL scores. The best results are in boldface.}
\vspace*{0.1in}
\scalebox{1}{
\begin{tabular}{c|c|c|c|ccc|ccc}
\hline
\hline
\multirow{2}{*}{\makecell[c]{Attention\\Pattern}} 
& \multirow{2}{*}{\makecell[c]{Block\\Size}} 
& \multirow{2}{*}{$w$} 
& \multirow{2}{*}{ $d$} 
& \multicolumn{3}{c|}{IRM} 
& \multicolumn{3}{c}{PSM}  \\
\cline{5-10}
&    &   &   & CSIG & CBAK & COVL & CSIG & CBAK & COVL \\
\hline
\hline
Noisy & -- & -- & -- & 2.26  & 1.80  & 1.67  & -- & -- & --  \\
\hline
\makecell[c]{Full Attention}    
& -- & -- & -- & 3.10 & 2.45 & 2.37     
               & 3.15 & 2.54 & 2.45 \\
\hline
Blockwise & \multirow{2}{*}{50} & \multirow{2}{*}{--} & \multirow{2}{*}{--}
& 3.08 & 2.42 & 2.34     
& 3.14 & 2.50 & 2.43 \\
\cline{1-1}
SepFormer  
&  &  & 
& 3.15 & 2.47 & 2.41     
& 3.19 & 2.55 & 2.49 \\
\hline
\multirow{3}{*}{\makecell[c]{Ripple}}     
   & \multirow{3}{*}{--} 
   & \multirow{3}{*}{12}
   & 16 
   &  \textbf{3.22} & \textbf{2.52} & \textbf{2.49}
   &  \textbf{3.28} & 2.60 & 2.58 \\
   &  &  & 24 
   & 3.21 & \textbf{2.52} & 2.47           
   & \textbf{3.28} & \textbf{2.61} & \textbf{2.59} \\
       
   &  &  & 50 
   & 3.20 & 2.50 & 2.47                                      
   & 3.25 & 2.59 & 2.56 \\
\hline
\hline

\end{tabular}}
\label{composite}
\vspace{-0.5em}
\end{table}
\textcolor{black}{Tables \ref{pesq} and \ref{estoi} report the comparison results of PESQ and ESTOI scores for the five SNR levels, respectively. Among the two training objectives, overall, PSM shows better performance than IRM. Our ripple self-attention patterns consistently improve over the unprocessed noisy recordings in terms of PESQ and ESTOI under all SNR levels. Taking the 5 dB SNR as a case, the ripple attention ($w\!=\!12$, $d\!=\!24$) with IRM provides 0.89 and 18.68\% gains on PESQ and ESTOI, respectively. It is also clear to observe that our ripple attention significantly improves PESQ and ESTOI scores of the full self-attention across all SNR conditions. In the case of 5 dB SNR, our sparse attention ($w\!=\!16$, $d\!=\!24$) with PSM improves full self-attention by 0.15 in PESQ and 2.36\% in ESTOI. Additionally, our attention consistently outperforms blockwise attention and SepFormer, which further confirms the superiority of the sparse attention design. For instance, our attention ($w=12$, $d=24$) with PSM provides 0.13 and 0.11 PESQ gains and 2.45\% and 2.06\% ESTOI gains over blockwise attention and SepFormer, respectively, for 5 dB SNR case. Table \ref{composite} reports the comparison results of CSIG, CBAK, and COVL scores averaged across all SNR levels, showing performance trends similar to the results in Tables \ref{pesq} and \ref{estoi}. On average, our ripple attention ($w\!=\!12$, $d\!=\!50$) with PSM improves CSIG by 0.10, 0.12, and 0.05, CBAK by 0.05, 0.08, and 0.03, and COVL by 0.10, 0.13, and 0.06 over full attention, blockwise attention, and SepFormer, respectively.}

\begin{figure}[!hbpt]
\centering
\includegraphics[width=0.60\columnwidth]{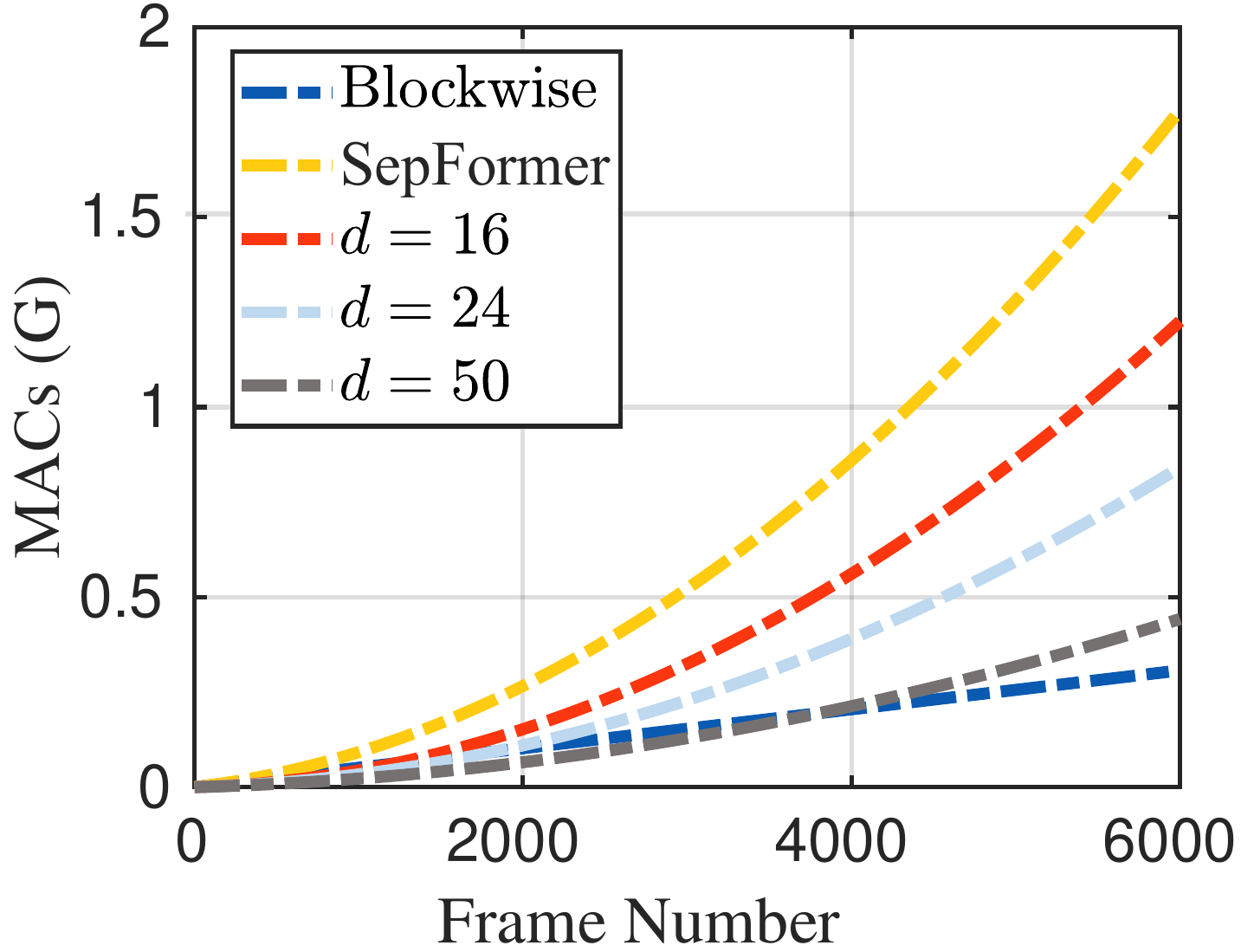}
\caption{The comparison (theoretical value) of multiply–accumulate operations (MACs) at various sequence lengths.}
\label{complexity}
\vspace{-1.7em}
\end{figure}

\textcolor{black}{In Fig. \ref{complexity}, we \textcolor{black}{theoretically} analyze the computation costs required by different attention patterns at various sequence lengths. The computation cost is measured in terms of multiply-accumulate operations (MACs). Among the three sparse self-attention patterns, the blockwise attention shows lower MACs. In addition, we can also find that our proposed attention patterns consistently exhibit lower MACs than SepFormer. Taking the performance and computation cost into account, our proposed ripple sparse attention provides a better trade-off than blockwise attention and SepFormer.}


\section{CONCLUSION}
\label{sec:5}
In this study, we propose a novel sparse self-attention for speech enhancement, termed ripple attention, which comprises a local band attention and a dilated attention. 
Local attention captures strong local correlations of speech signals at fine granularity, and the dilated attention allows the network model to capture global dependencies at coarse granularity. Our ripple sparse attention mechanism is investigated under different configurations of window size and dilation rate. Extensive speech enhancement experiments on two training objectives (IRM and PSM) show that our ripple attention consistently outperforms standard full self-attention, blockwise attention, and SepFormer in all cases, confirming the superiority of the design of our sparse attention. 

\vfill\pagebreak

\bibliographystyle{IEEEtran}
\bibliography{refs}

\begin{thebibliography}{10}
\providecommand{\url}[1]{#1}
\csname url@samestyle\endcsname
\providecommand{\newblock}{\relax}
\providecommand{\bibinfo}[2]{#2}
\providecommand{\BIBentrySTDinterwordspacing}{\spaceskip=0pt\relax}
\providecommand{\BIBentryALTinterwordstretchfactor}{4}
\providecommand{\BIBentryALTinterwordspacing}{\spaceskip=\fontdimen2\font plus
\BIBentryALTinterwordstretchfactor\fontdimen3\font minus
  \fontdimen4\font\relax}
\providecommand{\BIBforeignlanguage}[2]{{%
\expandafter\ifx\csname l@#1\endcsname\relax
\typeout{** WARNING: IEEEtran.bst: No hyphenation pattern has been}%
\typeout{** loaded for the language `#1'. Using the pattern for}%
\typeout{** the default language instead.}%
\else
\language=\csname l@#1\endcsname
\fi
#2}}
\providecommand{\BIBdecl}{\relax}
\BIBdecl

\bibitem{mmse}
Y.~Ephraim and D.~Malah, ``{Speech Enhancement Using a Minimum Mean-Square
  Error Short-Time Spectral Amplitude Estimator},'' \emph{IEEE Trans. Acoust.,
  Speech, Signal Process.}, vol. ASSP-32, no.~6, pp. 1109--1121, Dec. 1984.

\bibitem{gerkman2013}
T.~Gerkmann and M.~Krawczyk, ``{MMSE-Optimal Spectral Amplitude Estimation
  Given the STFT-Phase},'' \emph{{IEEE} Signal Process. Lett.}, vol.~20, no.~2,
  pp. 129--132, 2013.

\bibitem{zhang2019}
Q.~Zhang, M.~Wang, Y.~Lu, L.~Zhang, and M.~Idrees, ``A novel fast nonstationary
  noise tracking approach based on mmse spectral power estimator,''
  \emph{Digital Signal Processing}, vol.~88, pp. 41--52, 2019.

\bibitem{8740919}
Q.~Zhang, M.~Wang, Y.~Lu, M.~Idrees, and L.~Zhang, ``Fast nonstationary noise
  tracking based on log-spectral power mmse estimator and temporal recursive
  averaging,'' \emph{IEEE Access}, vol.~7, pp. 80\,985--80\,999, 2019.

\bibitem{wang2018supervised}
D.~Wang and J.~Chen, ``Supervised speech separation based on deep learning: An
  overview,'' \emph{IEEE/ACM Trans. Audio, Speech, Lang. Process.}, vol.~26,
  no.~10, pp. 1702--1726, 2018.

\bibitem{wang2014training}
Y.~Wang, A.~Narayanan, and D.~Wang, ``On training targets for supervised speech
  separation,'' \emph{IEEE/ACM Trans. Audio, speech, Lang. Process.}, vol.~22,
  no.~12, pp. 1849--1858, 2014.

\bibitem{williamson2015complex}
D.~S. Williamson, Y.~Wang, and D.~Wang, ``Complex ratio masking for monaural
  speech separation,'' \emph{IEEE/ACM Trans. Audio, speech, Lang. Process.},
  vol.~24, no.~3, pp. 483--492, 2015.

\bibitem{erdogan2015phase}
H.~Erdogan, J.~R. Hershey, S.~Watanabe, and J.~Le~Roux, ``Phase-sensitive and
  recognition-boosted speech separation using deep recurrent neural networks,''
  in \emph{Proc. ICASSP}, 2015, pp. 708--712.

\bibitem{SEGAN}
S.~Pascual, A.~Bonafonte, and J.~Serr{\`a}, ``{SEGAN}: Speech enhancement
  generative adversarial network,'' \emph{Proc. INTERSPEECH}, pp. 3642--3646,
  2017.

\bibitem{realse}
A.~Defossez, G.~Synnaeve, and Y.~Adi, ``Real time speech enhancement in the
  waveform domain,'' in \emph{Interspeech}, 2020.

\bibitem{weninger2015speech}
F.~Weninger, H.~Erdogan, S.~Watanabe, E.~Vincent, J.~L. Roux, J.~R. Hershey,
  and B.~Schuller, ``Speech enhancement with lstm recurrent neural networks and
  its application to noise-robust asr,'' in \emph{LVA/ICA}.\hskip 1em plus
  0.5em minus 0.4em\relax Springer, 2015, pp. 91--99.

\bibitem{TCN2018}
S.~Bai, J.~Z. Kolter, and V.~Koltun, ``An empirical evaluation of generic
  convolutional and recurrent networks for sequence modeling,'' \emph{arXiv
  preprint arXiv:1803.01271}, 2018.

\bibitem{deepmmse}
Q.~Zhang, A.~Nicolson, M.~Wang, K.~K. Paliwal, and C.~Wang, ``{DeepMMSE}: A
  deep learning approach to mmse-based noise power spectral density
  estimation,'' \emph{IEEE/ACM Trans. Audio, speech, Lang. Process.}, vol.~28,
  pp. 1404--1415, 2020.

\bibitem{GRN}
K.~Tan, J.~Chen, and D.~Wang, ``Gated residual networks with dilated
  convolutions for monaural speech enhancement,'' \emph{IEEE/ACM Trans. Audio,
  speech, Lang. Process.}, vol.~27, no.~1, pp. 189--198, 2018.

\bibitem{zhang2021temporal}
Q.~Zhang, Q.~Song, A.~Nicolson, T.~Lan, and H.~Li, ``Temporal convolutional
  network with frequency dimension adaptive attention for speech enhancement,''
  \emph{Proc. Interspeech}, pp. 166--170, 2021.

\bibitem{TFA}
Q.~Zhang, Q.~Song, Z.~Ni, A.~Nicolson, and H.~Li, ``Time-frequency attention
  for monaural speech enhancement,'' in \emph{Proc. ICASSP}, 2022, pp.
  7852--7856.

\bibitem{transformer}
A.~Vaswani, N.~Shazeer, N.~Parmar, J.~Uszkoreit, L.~Jones, A.~N. Gomez,
  {\L}.~Kaiser, and I.~Polosukhin, ``Attention is all you need,'' in
  \emph{NeurIPS}, 2017, pp. 5998--6008.

\bibitem{mhanet}
A.~Nicolson and K.~K. Paliwal, ``Masked multi-head self-attention for causal
  speech enhancement,'' \emph{Speech Communication}, vol. 125, pp. 80--96,
  2020.

\bibitem{tgsa}
J.~Kim, M.~El-Khamy, and J.~Lee, ``{T-GSA}: Transformer with gaussian-weighted
  self-attention for speech enhancement,'' in \emph{Proc. ICASSP}, 2020, pp.
  6649--6653.

\bibitem{sepformerstft}
D.~de~Oliveira, T.~Peer, and T.~Gerkmann, ``Efficient transformer-based speech
  enhancement using long frames and {STFT} magnitudes,'' in \emph{Proc.
  INTERSPEECH}, 2022, pp. 2948--2952.

\bibitem{tfaj}
Q.~Zhang, X.~Qian, Z.~Ni, A.~Nicolson, E.~Ambikairajah, and H.~Li, ``A
  time-frequency attention module for neural speech enhancement,''
  \emph{IEEE/ACM Transactions on Audio, Speech, and Language Processing},
  vol.~31, pp. 462--475, 2023.

\bibitem{sepformer}
C.~Subakan, M.~Ravanelli, S.~Cornell, M.~Bronzi, and J.~Zhong, ``Attention is
  all you need in speech separation,'' in \emph{Proc. ICASSP}, 2021, pp.
  21--25.

\bibitem{longformer}
I.~Beltagy, M.~E. Peters, and A.~Cohan, ``Longformer: The long-document
  transformer,'' \emph{arXiv preprint arXiv:2004.05150}, 2020.

\bibitem{panayotov2015librispeech}
V.~Panayotov, G.~Chen, D.~Povey, and S.~Khudanpur, ``Librispeech: an asr corpus
  based on public domain audio books,'' in \emph{Proc. ICASSP}, 2015, pp.
  5206--5210.

\bibitem{snyder2015musan}
D.~Snyder, G.~Chen, and D.~Povey, ``{MUSAN}: A music, speech, and noise
  corpus,'' \emph{arXiv preprint arXiv:1510.08484}, 2015.

\bibitem{hu2010tandem}
G.~Hu and D.~Wang, ``A tandem algorithm for pitch estimation and voiced speech
  segregation,'' \emph{IEEE Transactions on Audio, Speech, and Language
  Processing}, vol.~18, no.~8, pp. 2067--2079, 2010.

\bibitem{dean2010qut}
D.~B. Dean, S.~Sridharan, R.~J. Vogt, and M.~W. Mason, ``The {QUT-NOISE-TIMIT}
  corpus for the evaluation of voice activity detection algorithms,'' in
  \emph{Proc. INTERSPEECH}, 2010.

\bibitem{saki2016smartphone}
F.~Saki, A.~Sehgal, I.~Panahi, and N.~Kehtarnavaz, ``Smartphone-based real-time
  classification of noise signals using subband features and random forest
  classifier,'' in \emph{Proc. ICASSP}, 2016, pp. 2204--2208.

\bibitem{saki2016automatic}
F.~Saki and N.~Kehtarnavaz, ``Automatic switching between noise classification
  and speech enhancement for hearing aid devices,'' in \emph{in Proc. EMBC},
  2016, pp. 736--739.

\bibitem{steeneken1988description}
H.~J. Steeneken and F.~W. Geurtsen, ``{Description of the RSG-10 noise
  database},'' \emph{report IZF}, vol.~3, p. 1988, 1988.

\bibitem{Urban}
J.~Salamon, C.~Jacoby, and J.~P. Bello, ``A dataset and taxonomy for urban
  sound research,'' in \emph{Proc. ACM-MM}, 2014, pp. 1041--1044.

\bibitem{restcnsa}
Y.~Zhao, D.~Wang, B.~Xu, and T.~Zhang, ``Monaural speech dereverberation using
  temporal convolutional networks with self attention,'' \emph{IEEE/ACM Trans.
  Audio, speech, Lang. Process.}, vol.~28, pp. 1598--1607, 2020.

\bibitem{blockwise}
J.~Qiu, H.~Ma, O.~Levy, W.-t. Yih, S.~Wang, and J.~Tang, ``Blockwise
  self-attention for long document understanding,'' in \emph{Findings of
  EMNLP}, 2020, pp. 2555--2565.

\bibitem{localmodeling}
B.~Yang, Z.~Tu, D.~F. Wong, F.~Meng, L.~S. Chao, and T.~Zhang, ``Modeling
  localness for self-attention networks,'' in \emph{EMNLP}, 2018.

\bibitem{tii}
H.~Zhu, Q.~Zhang, P.~Gao, and X.~Qian, ``Speech-oriented sparse attention
  denoising for voice user interface toward industry 5.0,'' \emph{IEEE
  Transactions on Industrial Informatics}, vol.~19, no.~2, pp. 2151--2160,
  2023.

\bibitem{recommendation2001perceptual}
I.-T. Recommendation, ``Perceptual evaluation of speech quality ({PESQ}): An
  objective method for end-to-end speech quality assessment of narrow-band
  telephone networks and speech codecs,'' \emph{Rec. ITU-T P. 862}, 2001.

\bibitem{jensen2016algorithm}
J.~Jensen and C.~H. Taal, ``An algorithm for predicting the intelligibility of
  speech masked by modulated noise maskers,'' \emph{IEEE/ACM Trans. Audio,
  speech, Lang. Process.}, vol.~24, no.~11, pp. 2009--2022, 2016.

\bibitem{hu2007evaluation}
Y.~Hu and P.~C. Loizou, ``Evaluation of objective quality measures for speech
  enhancement,'' \emph{IEEE Trans. Audio, Speech, Lang. process.}, vol.~16,
  no.~1, pp. 229--238, 2007.

\end{thebibliography}

\end{document}